\documentclass[12pt]{iopart}

\usepackage[hidelinks]{hyperref}

\newcommand{\eprint}[2][]{%
  \if\relax\detokenize{#1}\relax
    arXiv:#2%
  \else
    #1:#2%
  \fi
}

\usepackage[numbers,square,sort&compress]{natbib}
\usepackage{url}
\usepackage{microtype}
\usepackage{graphicx}
\usepackage{subcaption}
\usepackage{booktabs}
\usepackage{fancyhdr}
\usepackage{amsmath}
\usepackage{amssymb}
\usepackage{mathtools}
\usepackage{amsthm}
\usepackage{float}
\usepackage{color,xspace}
\usepackage{enumitem}
\usepackage{placeins}
\usepackage{multirow}

\setlist{nolistsep,leftmargin=*}

\newcommand{\gnn}{\textsc{Gnn}\xspace}

\newcommand{\gcn}{\textsc{Gcn}\xspace}
\newcommand{\MLP}{\textsc{Mlp}\xspace}
\newcommand{\CG}{\mathcal{G}}
\newcommand{\CV}{\mathcal{V}}
\newcommand{\CE}{\mathcal{E}}
\newcommand{\CF}{\mathcal{F}}
\newcommand{\CW}{\mathcal{W}}
\newcommand{\CS}{\mathcal{S}}

\newcommand{\CH}{\mathcal{H}}
\newcommand{\CN}{\mathcal{N}}
\newcommand{\CL}{\mathcal{L}}
\newcommand{\bx}{\mathbf{x}}
\newcommand{\bh}{\mathbf{h}}
\newcommand{\bz}{\mathbf{z}}
\newcommand{\bP}{\mathbf{P}}
\newcommand{\bff}{\mathbf{f}}
\newcommand{\lgK}{\ell}          

\begin{document}

\title{Local learnable coarsening of unstructured grids}

\author{
Alexander Ryabov$^{1*}$, 
Viacheslav Naumov$^1$, 
Sergei Shumilin$^1$, 
Nikolay Yavich$^1$, 
Sayan Ranu$^3$, 
N. M. Anoop Krishnan$^3$, 
Evgeny Burnaev$^{1,2}$ and 
Vladimir Vanovskiy$^1$
}

\address{
$^1$ Artificial Intelligence Center, Skolkovo Institute of Science and Technology, 121205 Moscow, Russia \\
$^2$ Learnable Intelligence Group, Autonomous Non-Profit Organization Artificial Intelligence Research Institute (AIRI), 105064 Moscow, Russia \\
$^3$ Yardi School of Artificial Intelligence, Indian Institute of Technology Delhi, Hauz Khas, New Delhi, India 110016 \\
$^*$ Author to whom any correspondence should be addressed.
}

\ead{riabov.al.al@gmail.com}
\vspace{10pt}

\pagestyle{fancy}
\fancyhf{}
\lhead{Local learnable coarsening of unstructured grids}

\begin{abstract}
Efficient simulations of physical systems described by partial differential equations (PDEs) require computational methods to adequately reduce computational resources without compromising the accuracy of the simulation. One such method is to coarsen the computational grid.  In the case of strongly heterogeneous reservoirs, the accuracy of a coarse flow simulation is affected by the choice of the location of coarse cells, and the manner in which they average permeability. Here, we propose a local coarsening approach for Voronoi grids. We define a number of ‘anchors’ on the fine grid and grow a region from each anchor using a breadth-first search algorithm.  The same graph neural network processes each region independently, using cell coordinates relative to its anchor. Cells in overlapping regions can be assigned softly to several anchors. These weights determine the sites of the coarse Voronoi grid.  We set coarse permeability and storage from the fine field and coarse geometry, and learn corrections to face transmissibilities. We train through an implicit finite-volume solver using fine-grid sensor pressures and well rates as targets. We reduce the number of cells in each of the square and hexagonal Brugge domains by about an order of magnitude. Across both domains and storage rules, the global baseline’s mean test sensor error was 3.1 to 5.4 times that of the local model.
\end{abstract}


\section{Introduction}
\label{sec:intro}

Numerical simulation of flow through heterogeneous porous media is the basis for managing oil and gas reservoirs and for studying groundwater flows and storage in the subsurface \cite{aziz1979petroleum,lie2019introduction}. History matching and optimization of well controls are examples of applications requiring large numbers of forward simulations. For these applications, the cost of a single forward simulation on a geologically representative grid is the main restriction. Traditionally, a coarser grid is used, and an average of the properties of the constituent fine grid cells is assigned to each coarse grid cell (upscaling) \cite{durlofsky1991numerical,renard1997calculating,wen1996upscaling,durlofsky2005upscaling,farmer2002upscaling}.

Results of simulations depend strongly on the positions of coarse grid cells and the method used to average the properties within each of those cells.

In recent work we considered the coarse grid as the variable to be optimized. Specifically, we represented the coarse grid by a Voronoi tessellation of points, and differentiated the tessellation and a finite volume solver, with respect to the points \cite{shumilinself,shumilin2024method}. In a later work, we trained a graph neural network to perform this coarsening, without the need to perform optimization for each new permeability field \cite{ryabov2026learnable}. Then we extended this work to upscale two phase flow fields using a fully implicit upwind scheme \cite{naumov2026learnable}. 

In the earlier global method, the network scored each free fine cell against every coarse cluster. In our present method we implement locality by restricting each anchor to a region grown by breadth-first search. A cell in an overlap can receive assignment weights from each of the anchors whose regions contain it. Therefore, the model can still decide where to place coarse cell boundaries. As long as region overlap is bounded, the number of candidate pairs is linear in the size of the fine grid (Sec.\ref{sec:cost}). We share the local network across all regions and use relative coordinates with respect to the anchor. To encode the position of a region within the domain, we use distances to wells and sensors. We evaluate the local coarsener on held-out permeability fields.
Locality may also help it generalise to unseen source and sensor
configurations, which remains to be tested (Sec.~\ref{sec:transfer}).

The main contributions are:
\begin{enumerate}
  \item a local coarsener that uses overlapping regions and soft assignments to learn a coarsened Voronoi grid,
  \item coarse-property rules combining geometric-mean permeability with a straight-through
        gradient, storage based on the low-permeability volume fraction, and learned face
        transmissibility multipliers,
  \item a multi-seed comparison with global assignment on the square and hexagonal Brugge domains. The local coarsener achieved lower mean test sensor error on both Brugge domains under both storage rules, including when both models used three graph attention layers.
\end{enumerate}

\section{Related work}
\label{sec:related}
\paragraph{Upscaling and grid coarsening in reservoir simulation.}
In effective-permeability upscaling, the fine coefficients within a coarse block are
replaced by a single value or tensor. Analytical averages (arithmetic, harmonic,
geometric, power means) are exact only for layered or simple media. Flow based
upscaling solves the local problem on the fine scale using prescribed boundary
conditions \cite{durlofsky1991numerical}. Reviews
\cite{renard1997calculating,wen1996upscaling,farmer2002upscaling} indicate that no
single averaging rule is uniformly accurate across high-contrast media, which is
the case for our data sets. Structured or geometrically derived coarse grids are
typically used. Flow-based coarsening groups fine cells using indicators derived
from a flow calculation \cite{hauge2012flow,durlofsky2005upscaling}. We learn the
coarse grid properties from the fine scale response.

\paragraph{Multiscale and Multigrid methods.}
In multiscale methods, the global solution is computed using basis functions defined
on a coarse grid and incorporating information at the fine scale
\cite{jenny2003multiscale,moyner2016multiscale,de2022algebraic}. Multigrid methods use
grids of multiple resolutions to accelerate the solution on the fine grid
\cite{Stuben2001,vanek1996algebraic,Yavich2012}.

\paragraph{Graph coarsening and pooling.}
Multilevel partitioners such as METIS coarsen graphs by heavy-edge matching
\cite{karypis1998fast,hendrickson1995multilevel,dhillon2007weighted}. Spectral graph reduction approximates selected spectral properties of the original graph \cite{loukas2019graph}, and differentiable
approaches adjust edge weights in the coarse graph \cite{cai2021graph}. In machine
learning, differentiable pooling learns a soft assignment to clusters
\cite{ying2018hierarchical,bianchi2020spectral,gao2019graph}. For example, DiffPool
uses an assignment matrix with $N\times k$ entries for $N$ nodes and $k$ clusters.
In the context of graph neural networks (GNNs), other methods for local processing
are also used for large graph scaling
\cite{chiang2019cluster,zeng2021decoupling,hamilton2017inductive}. Our method constructs a coarse Voronoi grid by using local assignments to represent
the response of the fine grid sensors.

\paragraph{Learned surrogates and differentiable simulation.}
Neural Operators \cite{kovachki2023neural,li2021fourier} and convolutional surrogates
for subsurface flow \cite{mo2019deep,tang2020deep,wen2022ufno} and mesh-based graph
networks \cite{pfaff2021learning,alet2019graph,battaglia2018relational} learn the
solution operator. Although fast, they may introduce discrepancies
\cite{azizzadenesheli2024neural}. Some works employ differentiable simulation within
a neural framework to learn the solution of a given physical model
\cite{holl2020learning,schoenholz2020jax,li2024fully}. In this work, we retain the
finite volume solver and learn the coarse grid and corrections to face
transmissibilities to represent the coarse model, which, like the fine model,
preserves mass and satisfies the well controls.

\section{Problem formulation}
\label{sec:problem}

\subsection{Task data: grid, coefficient field, sources and sensors}
\label{sec:data}
The domain $\Omega\subset\mathbb{R}^2$ is discretised by a fine grid, the Voronoi tessellation of $N$ cell
centres $\bx_i$ clipped to $\Omega$. Cells $i$ and $j$ are neighbours if they share a face of length
$e_{ij}>0$; $h_{ij}=|\bx_i-\bx_j|$ and $V_i$ is the cell area. A task on this grid is specified by the
following data.
\begin{itemize}
  \item The \emph{coefficient field}: one scalar value $K_i>0$ per cell. It carries the heterogeneity
        that the coarse model has to reproduce (the permeability in the flow problem of
        Sec.~\ref{sec:flow}).
  \item The \emph{sources} $\CW$, each located in one cell. A source either prescribes its term in the
        balance of its cell (e.g.\ a rate), or it is \emph{controlled}: it holds a control value
        $u^{\mathrm c}_w$ and exchanges with its cell according to the state there. The exchange of a
        controlled source is itself an output, the \emph{source response} $f_w^n$. The set $\CW$ may be
        empty; the dynamics is then driven by the initial state.
  \item The \emph{sensors} $\CS$: cells in which the state is observed. Sensors do not act on the
        system.
  \item The initial state, the time step $\tau$ and the number of steps $n_t$.
\end{itemize}
The outputs of interest are the state history at the sensors, $u_s^n$, $s\in\CS$, $n=1,\dots,n_t$, and
the responses $f_w^n$ of the controlled sources, if there are any.

\subsection{Discrete model}
\label{sec:model}
The fine and the coarse model have the same cell--face form. The state is one value $u_i$ per cell;
cell $i$ has a capacity $m_i$; neighbouring cells exchange through their common face, with a face
coefficient $T_{ij}=T_{ji}$ computed from the coefficients of the two cells and the geometry of the
face, and in the linear case considered here the exchange is $T_{ij}(u_j-u_i)$; every source is coupled
to its cell; and the states are advanced by a fixed time integrator.

Two groups of ingredients have to be distinguished. The formula for $T_{ij}$, the capacity, the
coupling of a source to its cell and the time integrator are given by the equation and its
discretisation; for the flow problem they are given in Sec.~\ref{sec:flow}. The coarsening produces the
coarse cells and, for them, the coefficients, capacities and face coefficients of the same model; the
time integrator is the same for the fine and the coarse model. From the equation the method requires
only this cell--face structure, i.e.\ that cells exchange only with their face neighbours.

The present implementation further assumes a linear model whose system matrix does not change in
time, one scalar coefficient field and a capacity proportional to the cell area. Another equation
requires its own face formula and time integrator; if its heterogeneity enters through several fields,
additional input channels of the network and averaging rules for them; and if its discretisation
needs a minimum resolution, this bounds the size of the coarse cells.

\subsection{Coarsening task}
\label{sec:task}
A coarsening maps a task to a coarse Voronoi grid with $k+|\CF|\ll N$ cells, where $\CF$ is the set of
\emph{fixed} cells, comprising the source and sensor cells and their neighbours in the fine graph.
The coarse grid retains their centres and coefficients. The coarsening also supplies the
coefficients, capacities and face coefficients of the coarse model. Without sources, $\CF$ consists
of the sensors and their neighbours. The coarse model is the model of
Sec.~\ref{sec:model} on this grid with the same time integrator; because sources and sensors are fixed
cells, the coarse sensor history $\hat u_s^n$ and the coarse source responses $\hat f_w^n$ are directly
comparable with the fine ones. The coarsening is learned from a training set of tasks and applied to
new tasks without retraining or any fine simulation. The number $k$ of free coarse cells is a
parameter of the method.

\subsection{Error measure}
For one task, the sensor error is
\begin{equation}
  \varepsilon = \frac{1}{|\CS|}\sum_{s\in\CS}\frac{\mathrm{RMS}_n(\hat u_s^n - u_s^n)}{\max_{s'\in\CS}\mathrm{RMS}_n(u_{s'}^n)},
  \qquad \mathrm{RMS}_n(a^n)=\Bigl(\frac{1}{n_t}\sum_{n=1}^{n_t}(a^n)^2\Bigr)^{1/2},
  \label{eq:error}
\end{equation}
i.e.\ the relative $L_2$ error of the sensor curve for one sensor; normalising by the largest reference
curve keeps a sensor with an almost zero response from dominating. The same relative error of the
responses of the controlled sources is an optional second training target (Sec.~\ref{sec:loss}). We
also report the absolute RMS error on the same dimensionless scale as the pressure, since the
relative error depends on the plateau of the curve.

\subsection{Instance: single-phase flow}
\label{sec:flow}
The experiments use single-phase slightly compressible flow in a two-dimensional domain
$\Omega\subset\mathbb{R}^2$ with no-flow outer boundary:
\begin{equation}
  c\,\frac{\partial p}{\partial t} - \nabla\cdot\bigl(K(\bx)\nabla p\bigr) = \sum_{w\in\CW} q_w\,\delta(\bx-\bx_w),
  \qquad p(\bx,0)=p_0(\bx),
  \label{eq:pde}
\end{equation}
where the state is the pressure, $u=p$, the coefficient field is a scalar (isotropic) permeability $K$,
$c$ is the storage coefficient and $q_w$ the rate of well $w$. The sources are wells controlled by
their bottom-hole pressure: the control value is $u^{\mathrm c}_w=p^{\mathrm{bh}}_w$, and the source response
is the well rate. We solve a dimensionless flow problem. On the fine grid we use $c=1$, so the storage
of a cell equals its area, $m_i=V_i$. We consider convex two-dimensional domains with scalar
isotropic permeability and no separate porosity field.

The face coefficient is the two-point flux approximation with the harmonic mean of the cell
permeabilities, the transmissibility
\begin{equation}
  T_{ij} = \frac{2K_iK_j}{K_i+K_j}\,\frac{e_{ij}}{h_{ij}} .
  \label{eq:tpfa}
\end{equation}
It is consistent for a linear operator in divergence form with a scalar isotropic coefficient on a grid
whose faces are orthogonal to the lines that connect the cell centres, which holds for Voronoi grids.
For a tensor coefficient, for nonlinear problems (e.g.\ multiphase flow, where the transmissibility is
multiplied by an upwind mobility) or for other operators the face coefficient is different.
With the implicit Euler scheme and time step $\tau$,
\begin{equation}
  m_i\,\frac{p_i^{n+1}-p_i^{n}}{\tau} = \sum_{j\in\CN(i)} T_{ij}\bigl(p_j^{n+1}-p_i^{n+1}\bigr)
  + \omega_i\bigl(p^{\mathrm{bh}}_{w(i)}-p_i^{n+1}\bigr),
  \label{eq:scheme}
\end{equation}
where $\omega_i=1$ in a well cell and $0$ elsewhere (unit well index). The capacity $m_i$ equals
the cell area $V_i$ on the fine grid and on the coarse grid with full storage.
Section~\ref{sec:coarse-properties} defines the coarse capacity for rock storage. The matrix of
(\ref{eq:scheme}) does not change in time; it is factorised once per simulation. The well rate is
$q_w^n = \omega_i(p^{\mathrm{bh}}_w - p^n_i)$ for the well cell $i$.

\section{Method}
\label{sec:method}

\begin{figure}[htb]
  \centering
  \includegraphics[width=\linewidth]{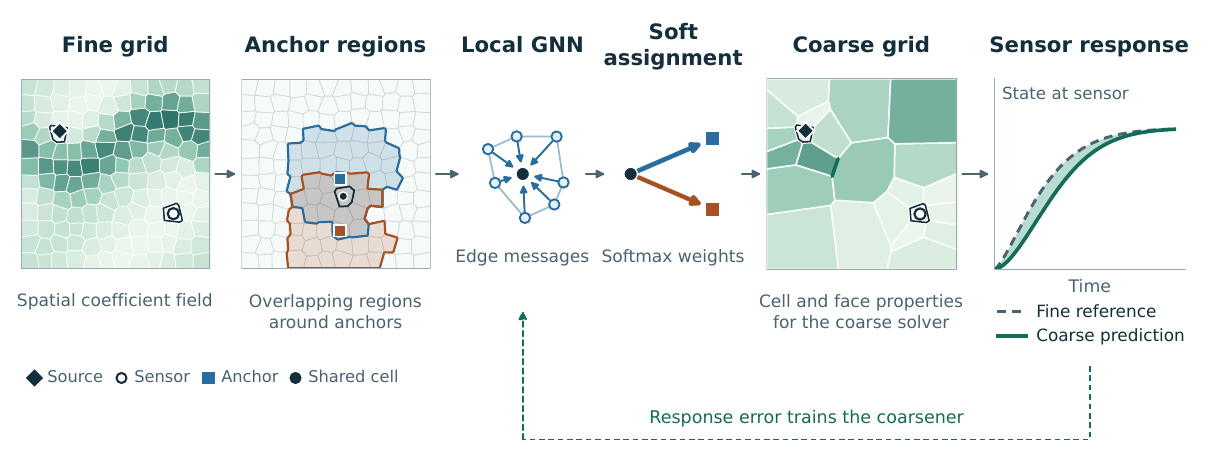}
  \caption{Local learnable coarsening. Overlapping regions define the candidate anchors for each
  free fine cell. The local \gnn processes each region separately. The assignment scores give
  soft weights for the candidate anchors, and these weights determine the coarse Voronoi sites.
  The coarse solver predicts the sensor pressure shown here and the well rate. During training,
  we compare both histories with their fine-grid references. The meshes and curves are illustrative.}
  \label{fig:overview}
\end{figure}

We prepare the fine graph, fixed cells, anchors and their regions once per task
(Figure~\ref{fig:overview}). The coarsener predicts local assignments, which determine the
coarse grid and its properties. We train through the coarse simulation using the fine-grid
outputs as targets. The stages after preprocessing are differentiable.

\FloatBarrier

\subsection{Fine graph and fixed cells}
\label{sec:fixed}
The fine graph $\CG=(\CV,\CE)$ has the cells as nodes and the Delaunay neighbours as edges, with
weights $T_{ij}$. Delaunay neighbours that touch only at a corner ($e_{ij}=0$, $T_{ij}=0$) remain edges of
$\CG$ for the breadth-first search but carry zero weight in the message passing (Sec.~\ref{sec:local}).
The fixed cells $\CF$ are the cells of the sources and sensors and their neighbours in $\CG$
(Sec.~\ref{sec:task}). We keep their centres and coefficients when building the coarse Voronoi grid.
Their boundaries and areas are recomputed in the coarse tessellation. All other cells are \emph{free}.

\subsection{Anchors (cluster centres)}
\label{sec:anchors}
The free coarse cells are built around $k$ anchor cells $a_1,\dots,a_k$, one per coarse cell. Two
placements are implemented:
\begin{itemize}
  \item \textbf{grid} (default): the points of a regular grid over the bounding box of the free cells,
        snapped to distinct free cells;
  \item \textbf{$k$-means}: the centres of $k$-means clustering \cite{lloyd1982least} of the free cells in the
        feature space $(x, y, \beta\,\tilde\lgK)$, with $\tilde\lgK$ the standardised $\log_{10}K$,
        snapped to distinct cells.
\end{itemize}
We use a regular grid in the main experiments so that the initial anchor placement
is independent of the random seed.
\ref{app:anchor-placement} compares grid and $k$-means anchors on both Brugge domains.
In both cases a fraction $\rho k$ of the anchors is moved next to every source or sensor whose
neighbourhood is heterogeneous, i.e.\ the standard deviation of $\log_{10}K$ over its $n_{\mathrm{nb}}$
nearest cells exceeds $s_{\mathrm{d}}$: walking outwards from the source or sensor, every
$m_{\mathrm{d}}$-th free cell becomes an anchor. The values of these and all other constants are given
in ~\ref{app:hyper}.
Densifying the anchors increases local resolution near sources and sensors, where a coarse cell
may otherwise average over a thin low-permeability barrier.
Anchors depend only on the task, not on the network, and are computed once.

\subsection{Local regions and local \gnn}
\label{sec:local}
\paragraph{Regions.} The region $\CH_j$ of anchor $a_j$ is the set of cells within $r$ graph hops of $a_j$,
i.e.\ the result of a breadth-first search of depth $r$ on $\CG$. The depth is
\begin{equation}
  r = \max\bigl(L+1,\ \lceil \kappa\, d_a/h\rceil\bigr),
\end{equation}
where $L$ is the number of \gnn layers, $d_a$ the median distance from an anchor to its nearest
anchor, $h$ the fine spacing and $\kappa$ a coverage factor; thus the regions cover the domain also
when the anchors are sparse. Regions of neighbouring anchors overlap; a free cell belongs
to all regions that contain it. A free cell outside all regions (rare) is assigned to its nearest
anchor.

As a variant (\textbf{hydraulic regions}), $\CH_j$ consists of the same number of cells taken in order
of the hydraulic distance from $a_j$: Dijkstra's distance on $\CG$ with edge cost
$1+\log_{10}(T_{\max}/T_{ij})$, so that a step costs at least one and crossing a face that is $m$ orders
of magnitude less conductive costs $1+m$. 

\paragraph{Cell features.} Every fine cell has the input vector
\begin{equation}
  \bff_i = \bigl(\log_{10}K_i,\ \pi_i,\ \chi^{\CS}_i,\ \chi^{\partial}_i,\ e^{-d^{\mathcal W}_i/\lambda},\ e^{-d^{\CS}_i/\lambda}\bigr),
\end{equation}
with $\pi_i$ the control value in the cell of a controlled source (the bottom-hole pressure in the flow
problem) and a small constant elsewhere, $\chi^{\CS}_i$ and $\chi^{\partial}_i$ indicators of sensor and
boundary cells, $d^{\mathcal W}_i$, $d^{\CS}_i$ the distances to the nearest source and sensor (without
sources, the source features are constant), and $\lambda$ a fixed multiple $c_\lambda$ of the median pairwise cell distance.
\emph{Absolute coordinates are not inputs.}

\paragraph{Local \gnn.} We process the regions in one batch of disconnected subgraphs. Region $j$
is the induced subgraph $\CG[\CH_j]$, with its nodes relabelled for batching. A cell that belongs to
several regions has a separate hidden state in each. The input of cell $i$ in region $j$ is
\begin{equation}
  \bh^{(0)}_{i|j} = \bigl[\bff_i,\ \bx_i-\bx_{a_j}\bigr].
\end{equation}
We apply $L$ graph convolution layers \cite{kipf2017semi} with ELU activations and edge weights
$w_{ij}=\log_{10}T_{ij}-\min\log_{10}T+1$ on faces and $0$ on corner-only neighbours. The coordinate
offsets are invariant to a joint translation of a cell and its anchor. The distance and boundary
features still locate the region relative to wells, sensors and the edge of the domain.

\subsection{Soft assignment and coarse points}
\label{sec:soft-assignment}
For every pair (free cell $i$, anchor $j$) with $i\in\CH_j$ the score is
\begin{equation}
  s_{ij} = \MLP\bigl(\bh_{i|j},\ \bh_{a_j|j},\ \bx_i-\bx_{a_j},\ \lgK_i-\lgK_{a_j}\bigr)
           - \gamma\,\frac{|\bx_i-\bx_{a_j}|^2}{2\sigma^2},
\end{equation}
with a one-hidden-layer \MLP (LeakyReLU), a Gaussian distance prior of strength $\gamma$ and width
$\sigma=c_\sigma\times$ the median pairwise anchor distance, and the membership is the softmax with
temperature $T_s$ over the anchors whose regions contain $i$,
\begin{equation}
  M_{ij} = \frac{\exp(s_{ij}/T_s)}{\sum_{j':\,i\in\CH_{j'}}\exp(s_{ij'}/T_s)} .
\end{equation}
The coarse point, coefficient and embedding of coarse cell $j$ are membership- and volume-weighted means,
\begin{equation}
  \bP_j=\frac{\sum_i M_{ij}V_i\bx_i}{\sum_i M_{ij}V_i},\quad
  \lgK^{\mathrm{soft}}_j=\frac{\sum_i M_{ij}V_i\lgK_i}{\sum_i M_{ij}V_i},\quad
  \bz_j=\frac{\sum_i M_{ij}V_i\bh_{i|j}}{\sum_i M_{ij}V_i},
\end{equation}
and $\bP_j$ is softly clipped to a disc around its anchor,
$\bP_j\leftarrow\bx_{a_j}+\boldsymbol\delta\,r_{\max}\tanh(|\boldsymbol\delta|/r_{\max})/|\boldsymbol\delta|$,
$\boldsymbol\delta=\bP_j-\bx_{a_j}$, with $r_{\max}=c_r\times$ the smallest anchor distance
(\textbf{global} radius, default) or $c_r\times$ the distance to the anchor's nearest neighbour
(\textbf{local} radius, variant). The clip keeps the coarse Voronoi grid valid.

\subsection{Coarse properties}
\label{sec:coarse-properties}
\paragraph{Coarse grid.} The coarse grid is the Voronoi tessellation of $\{\bP_j\}\cup\{\bx_i: i\in\CF\}$
clipped to $\Omega$, computed differentiably \cite{shumilin2024method}.

\paragraph{Coarse coefficient.} Two rules are implemented for $K_j$ (the permeability in the flow problem).
\textbf{Soft} (baseline): $K_j=10^{\lgK^{\mathrm{soft}}_j}$, the geometric mean over the soft members.
\textbf{Voronoi} (default): $K_j$ is the volume-weighted geometric mean of the fine cells whose centres
lie inside the coarse Voronoi cell $j$; the gradient is passed through a soft version of the
nearest-point assignment, $\mathrm{softmax}_j(-|\bx_i-\bP_j|^2/\tau_K)$ with
$\tau_K=c_\tau h\times$ the median nearest-coarse-point distance (straight-through estimator). A coarse
cell that contains no fine centre, and the fixed cells, keep the soft value. The Voronoi rule makes the
coefficient consistent with the geometry used by the solver.

\paragraph{Capacity (storage).} In \textbf{full} mode we set the capacity $m_j$ to the area
$V_j$ of the coarse Voronoi cell, as with $c=1$ on the fine grid. The coarse solver uses a single
pressure per cell, so the full area contributes to accumulation when that pressure changes.
On the fine grid, low-permeability parts of the same region may respond more slowly. We account
for this approximately in \textbf{rock} mode by using
$m_j=V_j\max(1-\phi^{\mathrm{b}}_j,\,\phi_{\min})$, where $\phi^{\mathrm{b}}_j$ is the
area-weighted fraction of fine cells with $\log_{10}K<\lgK_{\mathrm{b}}$ among those closest to
$\bP_j$. We hold the reduction factor fixed during each simulation and stop its gradient.
The rule is intended for the time window of the experiments ($t\leq 2$ on Brugge). In a connected
domain with no-flow boundaries, one well at constant bottom-hole pressure and uniform initial
pressure below that value, reducing total capacity also reduces the long-time cumulative
injection.

\paragraph{Edge head.} Averaging cannot represent sub-cell structures such as a thin barrier
that cuts a coarse face. A learned head multiplies every coarse face coefficient,
$T_{jl}\leftarrow T_{jl}\,10^{d_{jl}}$, with
\begin{equation}
  d_{jl} = \mathrm{clip}\bigl(\MLP(\bz_j+\bz_l,\ |\bz_j-\bz_l|,\ \lgK_j+\lgK_l,\ |\lgK_j-\lgK_l|,\ e_{jl},\ h_{jl},\ e_{jl}/h_{jl}),\ \pm d_{\max}\bigr),
\end{equation}
symmetric in $j,l$ ($\bz=0$ for fixed cells). The last layer is initialised at zero, so training
starts from the uncorrected face coefficient of the model ((\ref{eq:tpfa}) for flow).

\subsection{Coarse simulation and loss}
\label{sec:loss}
The coarse model, i.e.\ the time integrator of the fine model on the coarse grid ((\ref{eq:scheme})
for flow), is solved in float32 and is differentiable; the fine reference uses a sparse LU in float64
and is computed once per task. The loss of a task is
\begin{equation}
  \CL = \varepsilon_{\mathrm{sensor}} + \lambda_q\,\varepsilon_{\mathrm{src}},
\end{equation}
with $\varepsilon_{\mathrm{sensor}}$ from (\ref{eq:error}) and $\varepsilon_{\mathrm{src}}$ the same
relative error for the source responses. At prescribed bottom-hole pressure, the well rate tells us
how much fluid enters the domain. We compare the coarse and fine well-rate histories to supervise
this response alongside pressure at the sensor. We use $\lambda_q=1$ for every learned model in
the comparison. The source term is omitted when no source is controlled.

\subsection{Training}
We train with Adam, using separate learning rates for the coarsener and the edge head. We shuffle
the tasks into minibatches, clip the gradient norm and use early stopping
(\ref{app:hyper}). For each run we keep the checkpoint with the lowest validation sensor error,
using validation only for checkpoint selection and early stopping. The run is deterministic
for a given seed and environment.

\subsection{Summary of variants}
Table~\ref{tab:variants} lists the options for the coarsening method described here.

\begin{table}[htb]
\centering
\caption{Options of the implementation. Default in bold.}
\label{tab:variants}
\footnotesize
\setlength{\tabcolsep}{3pt}
\begin{tabular}{@{}lll@{}}
\toprule
component & options & section \\
\midrule
anchors & \textbf{grid} / $k$-means & \ref{sec:anchors} \\
regions & \textbf{$r$-hop BFS} / hydraulic (Dijkstra) & \ref{sec:local} \\
coarse $K$ & \textbf{Voronoi} / soft & \ref{sec:coarse-properties} \\
storage & \textbf{rock} / full & \ref{sec:coarse-properties} \\
edge head & \textbf{on} / off & \ref{sec:coarse-properties} \\
source-response target (well rates) & \textbf{on} / off & \ref{sec:loss} \\
clip radius & \textbf{global} / local & \\
prior width & \textbf{pairwise} / nearest-anchor & \\
input features & \textbf{standard} / original / clean & ~\ref{app:features} \\
corner-only edges in \gcn & \textbf{weight 0} / weight 1 & \\
number of free coarse cells $k$ & any & \\
assignment & \textbf{local} / global (previous method, for comparison) & \ref{sec:global-cmp} \\
\bottomrule
\end{tabular}
\end{table}

\section{Experiments}
\label{sec:experiments}

\subsection{Datasets}
\label{sec:datasets}

\subsubsection{Brugge.}
\label{sec:brugge}
The permeability fields are taken from the first Schelde layer of the Brugge reservoir model
\cite{peters2010brugge,brugge_dataset}. This fluvial layer was also used in the two-phase flow
study of Naumov et al.~\cite{naumov2026learnable}. The dataset contains 100 fields on the same fine
grid: the Voronoi tessellation of a $64\times64$ lattice in the unit square, $N=4096$ cells.
We use the permeability values stored in the dataset directly as the dimensionless coefficient $K$
in (\ref{eq:pde}). The values range from $7\cdot10^{-8}$ to $3.6$, spanning $7.7$ decades.
Each task has one source, an injector well controlled by its bottom-hole pressure $100$
(Sec.~\ref{sec:flow}), and one sensor cell; there are five distinct injector--sensor pairs, each used by 20 fields. $p_0=0$,
$\tau=0.02$, $n_t=100$. The reference curves are the fine solution of (\ref{eq:scheme}) in float64.
We shuffle the fields with a fixed seed and split them into 60 training, 20 validation and 20 test
fields. Coarse grids have $k=300$ free cells plus about 13 fixed cells; with this $k$ the region depth
is $r=3$. We use $\lgK_{\mathrm{b}}=-3$ as the permeability cutoff for rock storage and set
the source input feature $\pi_i$ to the bottom-hole pressure $100$.

\begin{figure}[htbp]
  \centering
  \includegraphics[width=0.92\linewidth]{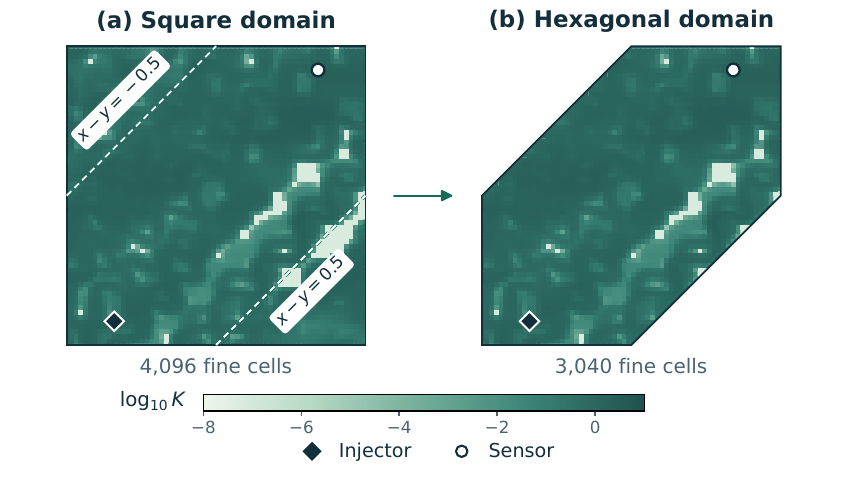}
  \caption{The same Brugge test field on the square and
  hexagonal fine grids. Colours show $\log_{10}K$ on a common scale. The dashed lines
  in (a) mark the cuts that define the domain in (b). The injector and sensor positions
  are the same in both domains.}
  \label{fig:brugge-domains}
\end{figure}

\paragraph{Hexagonal domain.} To see whether the results depend on the square domain, we also cut the
same 100 fields to the hexagon $|x-y|\le0.5$, i.e.\ remove the corners $(1,0)$ and $(0,1)$; all
injector--sensor pairs lie near the diagonal $x=y$ and stay inside. This leaves $N=3040$ cells, and the
reference curves are computed again with the fine solver. We use $k=223$ free coarse cells, which keeps
the ratio of fine to coarse cells of the square. Shuffling and split are the same as for the square, so
a field has the same number in both datasets. Figure~\ref{fig:brugge-domains}
shows one test field on both domains.

\FloatBarrier

\subsection{Protocol and metrics}
We train each learned configuration with three seeds. For the main comparisons we report the mean
$\pm$ standard deviation over seeds of the mean error over the training, validation and test
fields. Relative errors in the tables are given in per cent. Averages over all fields hide that a
minority of hard reservoirs dominates the error; we
therefore also report the mean over a fixed \emph{hard set}, the five test fields on which a
classical coarsening without learning (a $k$-means partition of the free cells with the geometric mean
of $K$) has the largest error; the set is chosen separately for each domain and is the same for all
models. We also report the mean of the five worst test fields of each model and the
absolute RMS error on the test fields. We select checkpoints by validation sensor error.
For each local--global pair, we average the errors across seeds for each of the 20 test fields
and apply the Wilcoxon signed-rank test to the paired differences. We ran each job using Python 3.11.11 and PyTorch 2.3.0.

\subsection{Comparison with the global coarsener}
\label{sec:global-cmp}
We compare with the global coarsener of \cite{ryabov2026learnable}. It applies three graph
attention layers to the entire fine graph. A linear layer and a softmax then give each free
cell a membership weight for every coarse cluster. For Brugge, we use $\log_{10}K$ and
$\log_{10}T$ as inputs because the coefficients span several orders of magnitude. We also
increase the hidden width from 3 to 16 to match the interface to the edge head, use
volume-weighted cluster means, and add sources and sensors to the fixed set
(Sec.~\ref{sec:fixed}).

On each domain, we train both coarseners with rock storage and then train both again with
full storage. Within each comparison, the models use the same fields, coarse-property rules,
edge head, solver and loss. We allow up to 400 epochs for the local model and 800 for the
global model, which converges more slowly, and use the same validation criterion for
checkpoint selection.

\begin{table}[htb]
\centering
\caption{Local and global assignment with rock and full storage. Values are means
$\pm$ sample standard deviations over three seeds. All error columns are in per cent
except absolute RMS, which uses the dimensionless pressure scale. The hard set is fixed
within each domain. The last column gives
the number of test fields on which the global model has lower seed-averaged error and
the paired Wilcoxon $p$-value.}
\label{tab:global-cmp}
\footnotesize
\setlength{\tabcolsep}{3pt}
\begin{tabular}{@{}lccccccc@{}}
\toprule
assignment & train & val & test & hard set & worst-5 & abs.\ RMS & global wins, $p$ \\
\midrule
\multicolumn{8}{@{}l}{\emph{Square, rock storage}} \\
  local & \(2.40{\pm}0.18\) & \(3.45{\pm}0.35\) & \(1.80{\pm}0.05\) & \(1.38{\pm}0.27\) & \(3.51{\pm}0.04\) & \(0.47{\pm}0.01\) & --- \\
  global & \(7.47{\pm}1.53\) & \(8.78{\pm}2.08\) & \(9.66{\pm}4.13\) & \(10.22{\pm}3.80\) & \(26.31{\pm}12.57\) & \(2.26{\pm}0.81\) & 1/20, \(3.8\cdot10^{-6}\) \\
\midrule
\multicolumn{8}{@{}l}{\emph{Square, full storage}} \\
  local & \(4.68{\pm}0.60\) & \(3.47{\pm}0.35\) & \(3.21{\pm}0.27\) & \(5.72{\pm}0.79\) & \(8.04{\pm}1.12\) & \(0.86{\pm}0.05\) & --- \\
  global & \(9.80{\pm}1.05\) & \(11.38{\pm}2.56\) & \(10.00{\pm}3.77\) & \(10.68{\pm}1.82\) & \(22.42{\pm}11.41\) & \(2.48{\pm}0.76\) & 1/20, \(2.1\cdot10^{-4}\) \\
\midrule
\multicolumn{8}{@{}l}{\emph{Hexagon, rock storage}} \\
  local & \(3.08{\pm}0.36\) & \(4.25{\pm}0.08\) & \(2.05{\pm}0.47\) & \(1.63{\pm}0.78\) & \(4.67{\pm}1.32\) & \(0.63{\pm}0.14\) & --- \\
  global & \(6.33{\pm}1.29\) & \(7.89{\pm}0.65\) & \(7.51{\pm}1.28\) & \(7.26{\pm}3.48\) & \(16.40{\pm}4.08\) & \(2.34{\pm}0.41\) & 1/20, \(5.7\cdot10^{-6}\) \\
\midrule
\multicolumn{8}{@{}l}{\emph{Hexagon, full storage}} \\
  local & \(3.24{\pm}1.01\) & \(4.54{\pm}0.27\) & \(2.53{\pm}0.51\) & \(3.22{\pm}0.68\) & \(5.65{\pm}1.61\) & \(0.80{\pm}0.16\) & --- \\
  global & \(7.71{\pm}1.28\) & \(9.22{\pm}2.79\) & \(9.58{\pm}1.21\) & \(12.46{\pm}0.48\) & \(20.51{\pm}1.66\) & \(3.05{\pm}0.31\) & 1/20, \(3.8\cdot10^{-6}\) \\
\bottomrule
\end{tabular}
\end{table}

The local model had lower sensor error on 19 of the 20 test fields in each comparison
(Table~\ref{tab:global-cmp}). With rock storage, the global mean test error was 5.4 times
the local error on the square and 3.7 times on the hexagon. With full storage, the ratios
were 3.1 and 3.8. The local model also had lower mean error on the fixed hard sets and
lower absolute RMS error in all four comparisons.

Rock storage gave lower mean test error than full storage for both coarseners on both domains.
The local model retained its advantage when we used the full capacity of each coarse cell.
\ref{app:assignment-mask} isolates the effect of the assignment mask in a shared
architecture on the square with full storage.
\ref{app:gnn-depth} compares two and three local \gcn layers on both domains.
The local coarsener also had lower mean test error in all four settings when both models
used three GAT layers (\ref{app:local-attention}).
Grid anchors gave lower mean test error in three of the four settings,
while $k$-means gave lower error on the square domain with full storage
(\ref{app:anchor-placement}).

\subsection{Transfer to new configurations}
\label{sec:transfer}
The experiments evaluate generalisation to held-out Brugge fields within each domain. Transfer to
unseen source and sensor configurations remains open.

\subsection{Computational cost}
\label{sec:cost}
Let $N$ be the number of fine cells and $k$ the number of free coarse cells. The local
assignment scores $O(N)$ cell--anchor pairs when the average number of regions covering a
cell is bounded. The global assignment scores $O(Nk)$ pairs.

The cost of the coarse simulation depends on the size of the coarse grid and the number of
time steps. On the fine grid, sparse factorisation costs $O(N^{3/2})$ in two dimensions under
suitable ordering.

\section{Conclusion}
\label{sec:conclusion}
Over Brugge domains, the local coarsener maintained a lower mean test sensor error than the global baseline under both storage rules. With full storage, the mean error of the global model was 3.1 times that of the local model on the square domain and 3.8 times that of the local model on the hexagon domain. Under bounded region overlap, the local assignment scores $O(N)$ cell--anchor pairs. In the controlled square comparison with full storage, the assignment mask reduced the mean test error from 5.18\% to 3.95\%, with differences between fields (\ref{app:assignment-mask}).

The cell--face formulation can also be used with other equations. We have yet to test this local
coarsener on larger or three-dimensional grids, in multiphase flow, or on well configurations
unseen during training.

\section*{Code availability}
The source package will be released through the
\href{https://github.com/AlexanderFreeman/graph-reduction-for-physics}{project repository on GitHub}.
It includes the coarsening code and the scripts for training, evaluation and baselines. It also
includes scripts for the comparisons in Table~\ref{tab:global-cmp} and the assignment-mask study in
\ref{app:assignment-mask}. Running the Brugge experiments requires a local copy of the dataset.
The Python and PyTorch versions used in the reported runs are given in Sec.~\ref{sec:experiments}.

\section*{Acknowledgements}
The study was supported by the Russian Science Foundation grant No.~24-41-02035,
\url{https://rscf.ru/en/project/24-41-02035/}

\paragraph{Use of generative AI.}
The authors used generative AI tools for language editing and stylistic polishing of the manuscript,
including checking grammar and punctuation, clarifying phrasing, and improving the flow of the text.
The authors take responsibility for the final manuscript.

\bibliography{refs}

\clearpage
\appendix

\section{Hyperparameters}
\label{app:hyper}
\begin{table}[htb]
\centering
\caption{Constants of the method and their values in the experiments. Values marked $^\ast$ depend
on the units or the grid of the dataset; the others are the same for all datasets.}
\label{tab:hyper}
\footnotesize
\setlength{\tabcolsep}{2.5pt}
\begin{tabular}{@{}lll@{}}
\toprule
symbol & meaning & value \\
\midrule
$k$ & free coarse cells & 300$^\ast$ \\
$\rho$ & densified anchors per heterogeneous source or sensor, fraction of $k$ & 0.1 \\
$n_{\mathrm{nb}}$, $s_{\mathrm{d}}$ & neighbourhood size and $\log_{10}K$ std threshold for densification & 40, 1.0 \\
$m_{\mathrm{d}}$ & stride of densified anchors & 6 \\
$\beta$ & weight of $\log_{10}K$ in $k$-means; $k$-means seed & 0.2; 40 \\
$L$ & local \gcn layers & 2 \\
$\kappa$ & region coverage factor & 0.75 \\
--- & hidden size, embedding size, assignment \MLP hidden size & 16, 16, 16 \\
$T_s$ & softmax temperature of the assignment & 0.4 \\
$\gamma$, $c_\sigma$ & strength and relative width of the distance prior & 5, 0.5 \\
$c_r$ & clip radius, fraction of the anchor distance & 0.45 \\
$c_\lambda$ & distance-feature scale, multiple of the median pairwise cell distance & 8 \\
$c_\tau$ & width of the Voronoi-$K$ gradient & 0.4 \\
$\lgK_{\mathrm{b}}$ & permeability cutoff for rock storage & $-3^\ast$ \\
$\phi_{\min}$ & storage floor & 0.02 \\
--- & edge head \MLP & $37\to32\to32\to1$, SiLU \\
$d_{\max}$ & bound of the $\log_{10}$ face multiplier & 3 \\
$\lambda_q$ & weight of the source-response error (target on) & 1 \\
--- & learning rates (coarsener / edge head) & $3\cdot10^{-3}$ / $10^{-3}$ \\
--- & minibatch, gradient-norm clip & 10 tasks, 1.0 \\
--- & maximum epochs, early-stopping patience & 400, 120 \\
\end{tabular}
\end{table}
\FloatBarrier

\section{Input-feature variants and implementation notes}
\label{app:features}
The experiments use the \texttt{standard} features described in Sec.~\ref{sec:local}. The
\texttt{clean} variant scales source control values by their maximum absolute value and uses
distances normalised by the diagonal of the fine-cell bounding box.

\clearpage
\section{Controlled comparison of local and unrestricted assignment}
\label{app:assignment-mask}
We run this comparison on the square Brugge domain with full storage. We encode the fine
graph with two GCN layers and reuse each cell's embedding across candidate anchors. We score each
cell--anchor pair with a scaled dot product of learned projections of the two embeddings
and add the distance prior of Sec.~\ref{sec:soft-assignment}. Both versions score every pair. The
unrestricted version normalises over all anchors. In the local version, the softmax
includes only anchors whose regions contain the cell, with the nearest anchor as a
fallback for any uncovered cell.

Both versions use the same inputs, anchor locations, coarse-property rules, edge head and
solver. We start each pair of runs from the same weights and train both versions on
the same 60 training fields, using the combined pressure and well-rate loss.
We use seeds 0, 1 and 2, allow up to 400 epochs per run and select the checkpoint
by sensor error on the same 20 validation fields.

\begin{table}[htb]
\centering
\caption{Effect of restricting anchor assignments on the square Brugge domain with
full storage. Validation and test errors are percentages. We report the mean and
sample standard deviation over three seeds. Candidates are averaged over free cells
on the test fields.}
\label{tab:assignment-mask}
\begin{tabular}{@{}lcccc@{}}
\toprule
assignment & candidates per cell & validation & test & abs.\ RMS \\
\midrule
all anchors & 300 & \(4.37{\pm}0.39\) & \(5.18{\pm}1.65\) & \(1.31{\pm}0.41\) \\
local mask & 2.46 & \(3.58{\pm}0.25\) & \(3.95{\pm}0.61\) & \(1.01{\pm}0.17\) \\
\bottomrule
\end{tabular}
\end{table}

The local mask lowered mean test sensor error from 5.18 to 3.95\,\%.
The error was lower in 2 of the 3 matched runs and on 13 of the 20 test fields
after we averaged errors across seeds. The two-sided paired Wilcoxon test gave
$p=0.123$, so the evidence for an improvement across test fields is inconclusive.

\clearpage
\section{Depth of the local graph network}
\label{app:gnn-depth}
We test a third \gcn layer in the local coarsener of Sec.~\ref{sec:method}. We keep the
anchor regions and the assignment rule fixed, and initialise the parameters shared by the
two models identically for each seed. We train on both Brugge domains with each storage rule,
using the same data splits, loss and validation criterion as in Sec.~\ref{sec:global-cmp}.

\begin{table}[htb]
\centering
\caption{Test sensor error in per cent for the local coarsener with two or three \gcn layers.
Errors are means $\pm$ sample standard deviations over three seeds. The last column gives
the number of fields on which the three-layer model has lower seed-averaged error and
the two-sided paired Wilcoxon $p$-value, unadjusted for the four comparisons.}
\label{tab:gnn-depth}
\begin{tabular}{@{}llccc@{}}
\toprule
domain & storage & two layers & three layers & three-layer wins, $p$ \\
\midrule
Square & rock & \(1.80{\pm}0.05\) & \(3.61{\pm}1.29\) & 7/20, \(0.0266\) \\
Square & full & \(3.21{\pm}0.27\) & \(2.25{\pm}0.11\) & 11/20, \(0.349\) \\
Hexagon & rock & \(2.05{\pm}0.47\) & \(1.86{\pm}0.03\) & 12/20, \(0.674\) \\
Hexagon & full & \(2.53{\pm}0.51\) & \(2.43{\pm}0.40\) & 11/20, \(0.596\) \\
\bottomrule
\end{tabular}
\end{table}

The effect of the third layer depended on the storage rule. On the square, it raised
the test error in all three runs with rock storage and lowered it in all three with
full storage. On the hexagon, the mean changed by less than 0.2 percentage points under
either rule. We keep the two-layer network for the main comparison in Table~\ref{tab:global-cmp}.

\clearpage
\section{Three-layer local and global coarseners}
\label{app:local-attention}
We replace the three local \gcn layers of \ref{app:gnn-depth} with three graph attention
layers. Each layer uses one attention head, with width 16 and the same transformed
transmissibilities as edge features. We keep the anchor regions, input features and soft
assignment of the local coarsener. Faces with positive transmissibility carry messages.
We train each model on the same fields and with the same loss and validation criterion as
in \ref{app:gnn-depth}.

Table~\ref{tab:local-attention} includes the global coarsener from Table~\ref{tab:global-cmp}.
The two GAT coarseners use the same number and type of graph layers. They retain their
respective input features and assignment rules.

\begin{table}[htb]
\centering
\caption{Mean test sensor errors in per cent for the three-layer models. We report means
$\pm$ sample standard deviations over three seeds.}
\label{tab:local-attention}
\begin{tabular}{@{}llccc@{}}
\toprule
domain & storage & local GCN & local GAT & global GAT \\
\midrule
Square & rock & \(3.61{\pm}1.29\) & \(2.76{\pm}0.48\) & \(9.66{\pm}4.13\) \\
Square & full & \(2.25{\pm}0.11\) & \(2.04{\pm}0.36\) & \(10.00{\pm}3.77\) \\
Hexagon & rock & \(1.86{\pm}0.03\) & \(1.96{\pm}0.24\) & \(7.51{\pm}1.28\) \\
Hexagon & full & \(2.43{\pm}0.40\) & \(2.46{\pm}0.48\) & \(9.58{\pm}1.21\) \\
\bottomrule
\end{tabular}
\end{table}

The local GAT gave lower mean test error than the global GAT under both storage
rules on both domains. The global error was 3.5 to 4.9 times as large, and the local GAT was
more accurate on 17--20 of the 20 test fields after averaging over the seeds. Within the
local coarsener, switching from GCN to GAT lowered the mean error on the square.
On the hexagon, the changes were about 0.1 percentage points or less.

\clearpage
\section{Effect of anchor placement}
\label{app:anchor-placement}
We train the two-layer local GCN with grid and $k$-means anchors on both Brugge domains
under both storage rules. The two placements use the same number of anchors and the same
rule for densifying anchors near sources and sensors (Sec.~\ref{sec:anchors}).
We use the field splits, loss and checkpoint selection of Sec.~\ref{sec:global-cmp}.
We match the training seeds (0, 1 and 2) across placements and fix the $k$-means
clustering seed at 40.

\begin{table}[htb]
\centering
\caption{Mean sensor error on the 20 test fields for grid and $k$-means anchors, in per cent.
Values are means $\pm$ sample standard deviations over three training seeds. The last
column gives the number of fields on which $k$-means has lower seed-averaged error and
the two-sided paired Wilcoxon $p$-value, unadjusted for the four comparisons.}
\label{tab:anchor-placement}
\begin{tabular}{@{}llccc@{}}
\toprule
domain & storage & grid & $k$-means & $k$-means wins, $p$ \\
\midrule
Square & rock & \(1.80{\pm}0.05\) & \(2.54{\pm}0.25\) & 8/20, \(0.216\) \\
Square & full & \(3.21{\pm}0.27\) & \(2.73{\pm}0.26\) & 13/20, \(0.648\) \\
Hexagon & rock & \(2.05{\pm}0.47\) & \(3.07{\pm}0.61\) & 7/20, \(0.076\) \\
Hexagon & full & \(2.53{\pm}0.51\) & \(2.78{\pm}0.17\) & 11/20, \(0.985\) \\
\bottomrule
\end{tabular}
\end{table}

With rock storage, grid anchors gave lower test error for all three training seeds on
both domains. With full storage, $k$-means lowered the mean test error on the square
from 3.21 to 2.73\,\%. The paired tests gave limited evidence for a difference across fields
(Table~\ref{tab:anchor-placement}).

\end{document}